\documentclass{an}
\usepackage{graphicx}
\usepackage{times}
\usepackage{fancyhdr}
\begin{document}

\title{Forced Oscillations in Fluid Tori and Quasi--Periodic Oscillations}

\author{William H. Lee}

\institute{Instituto de Astronom\'{\i}a, UNAM, Apdo. Postal 70--264,
Cd. Universitaria, M\'{e}xico D.F. 04510}

\date{Received; accepted; published online}

\abstract{The kilo-Hertz Quasi--Periodic Oscillations in X--ray
binaries could originate within the accretion flow, and be a signature
of non--linear fluid oscillations and mode coupling in strong
gravity. The possibility to decipher these systems will impact our
knowledge of fundamental parameters such as the neutron star mass,
radius, and spin. Thus they offer the possibility to constrain the
nuclear equation of state and the rotation parameter of stellar--mass
black holes. We review the general properties of these oscillations
from a hydrodynamical point of view, when the accretion flow is
subject to external perturbations and summarize recent results.
\keywords{accretion disks --- gravitation --- hydrodynamics --- stars:
neutron --- X--rays: binaries}}

\correspondence{wlee@astroscu.unam.mx}

\maketitle

\section{Introduction}\label{intro}

The modulations of the X--ray flux in Low--Mass X--ray Binaries
(LMXBs) and their observed range in frequencies, from $\approx 300$~Hz
to $\approx 1200$~Hz (see van der Klis 2000; McClintock \& Remillard
2006; van der Klis, these proceedings, for reviews) strongly suggest
that dynamical time scales in the inner regions of the accretion disks
occurring in these systems are being probed (i.e., at orbits only a
few gravitational radii in circumference).

In Black Hole X--ray Binaries (BHXRBs), the high frequency
Quasi--Periodic Oscillations (QPOs) in the X--ray flux seem to be
fairly stable, exhibiting nearly constant frequencies. Additionally,
in the four galactic microquasars know to show more than one peak,
their frequencies are in a 3:2 ratio, strongly suggesting that a
resonance is responsible for their production (Klu\'{z}niak \&
Abramowicz 2000).

In Neutron Star (NS) systems, the twin peaks observed in the frequency
spectrum drift over a considerable interval (a few hundred Hz, as
mentioned above), yet are linearly correlated throughout this
range. In many sources the peak separation, while not constant, is
consistent with being half the burst oscillation frequency, and in
other sources with the burst frequency directly. In the one observed
instance where burst oscillations, the spin frequency of the pulsar,
and twin QPO peaks in the kHz range have been observed in the same
source (SAX J1808.4-3658), the burst oscillation {\em is} the spin
frequency, and is twice the peak separation, i.e.,
$\nu_{burst}=\nu_{s}=2 \Delta \nu$(QPO) (Wijnands et al. 2003). In
XTEJ1807--294, the peak separation is consistent with the spin
frequency (Linares et al. 2005), and there generally appears to be a
trend whereby the "slow" neutron stars have twin peaks with $\Delta
\nu \approx \nu_{s}$ and the "fast" ones show $\Delta \nu \approx
\nu_{s}$ (with the transition at about 300~Hz).

We have previously suggested (Klu\'{z}niak et al. 2004; Lee,
Abramowicz \& Klu\'{z}niak 2004) that the peculiar behavior of SAX
J1808 can be understood in terms of a nonlinear response of the
accretion disk, when subject to an external perturbation at a fixed
frequency. While it is currently not possible to study the detailed
structure and dynamical behavior of the accretion disk and its modes
of oscillations in full detail, general physical principles can still
be applied, and yield suggestive results. We have proposed that within
the accretion disk, as in other nonlinear systems, a subharmonic
response will appear in the presence of a perturbation, as will higher
harmonics of the perturbation itself. Specifically, a second QPO
frequency is to appear when two possible modes of oscillation are
separated by half the perturbing frequency, and thus couple.

This presentation is devoted to a numerical study of the non--linear
oscillations in such systems, by considering bound fluid tori
representing a portion of the accretion flow around the compact object
(in particular thin tori). We believe strong gravity plays an
important role in their behavior, and consider its effects in a
simplified approach. This is a purely hydrodynamical study, and so the
potentially important effects of the magnetic field have not been
considered for now.

\section{Hydrostatic equilibrium for fluid tori}\label{hydroeq}

Astrophysical fluid bodies often appear in two topologically different
varieties: spheres and tori. In the former, support against
gravitational collapse is mostly provided by pressure from within
(e.g., by photons, degenerate or ideal gas particles, neutrinos), with
a possible centrifugal contribution, while in the latter, it comes
mostly from rotation, with pressure providing corrections from simple
test particle motion. Each of these terms: hydrodynamical and
mechanical in nature, play an important role in the behavior of the
system under external perturbations, and on the propagation of waves
within such systems. As we will consider here the behavior and
possible observational consequences of accretion disks in the
particular context of LMXBs, we will focus mainly on toroidal
configurations. Nevertheless, knowledge gathered from quasi--spherical
systems is relevant, and indeed quite useful for interpreting these
systems. 

\subsection{Gravitational potentials}\label{potentials}

At this point we may consider various forms for the gravitational
potential of the central mass, $M$. In the Newtonian regime, obviously
$\Phi= \Phi_{\rm N}=-G M/r$. Since we are interested in applications
to systems containing compact objects such as Neutron Stars and Black
Holes, it is useful to consider pseudo--Newtonian potentials of the
kind proposed originally by Paczy\'{n}ski and Wiita (1980), namely
$\Phi_{\rm PW}=GM/(r-r_{g})$, where $r_{g}=2GM/c^{2}$ is the
gravitational, or Schwarszchild, radius of the central mass. The
important feature of this class of potentials is that they reproduce
the existence of a marginally stable orbit, and that of capture orbits
for finite values of the orbital angular momentum, $\ell$, both unique
features of General Relativity. In addition to the original form, we
have also used extensively a new pseudo-potential, $\Phi_{\rm
KL}=GM[1-\exp(r_{ms}/r)]/r_{ms}$, where $r_{ms}=3r_{g}$ (Klu\'{z}niak
\& Lee 2002). The main advantage of this expression is related to the
epicyclic frequencies for test particle motion.

In Newtonian gravity, a test particle in circular motion with
frequency $\nu_{\phi}$ around a point mass, $M$, will perform small
epicyclic motions when perturbed. These can be radial or vertical, but
will always occur at frequencies $\nu_{r}=\nu_{z}=\nu_{\phi} = \Omega
/2 \pi$. This is just a re--statement of the well-known fact that the
Newtonian orbits of a point mass are closed curves. In addition, no
matter what the value of angular momentum a particle has, $\ell$, one
can always place it a a certain radius, $r_{\rm circ}=\ell^{2}/GM$
such that it will be in circular orbit.

In strong gravity this is no longer the case, as the central well in
the potential is so powerful that two qualitatively new effects
appear. First, capture orbits exist even at finite $\ell$. Second, in
the most simple case of the Schwarzschild metric (static, spherically
symmetric geometry), the three--fold degeneracy between orbital and
epicyclic frequencies is broken, such that $\nu_{r} <
\nu_{z}=\nu_{\phi}$, and $\nu_{r}^{2} < 0$ for $r < r_{\rm ms}$. Radial
oscillations are thus unstable inside $r_{\rm ms}$, and no circular
orbit is possible in that region.

The particular form of $\Phi_{\rm KL}$ is such that the ratio of
radial to vertical epicyclic frequencies as a function of radius,
$\nu_{r}/\nu_{z}$ is exactly equal to the value computed in full
General Relativity in the Schwarszchild metric. Specifically, we have 
\begin{equation}
\nu_{z}^{2}=\frac{1}{4 \pi^{2}} \frac{GM}{r^{3}} \exp(r_{\rm ms}/r),
\end{equation}
and
\begin{equation}
\nu_{r}^{2}=\frac{1}{4 \pi^{2}} \left(1-\frac{r_{\rm ms}}{r}\right)
\frac{GM}{r^{3}} \exp(r_{\rm ms}/r).
\end{equation}

\begin{figure}
\resizebox{\hsize}{!}
{\includegraphics[]{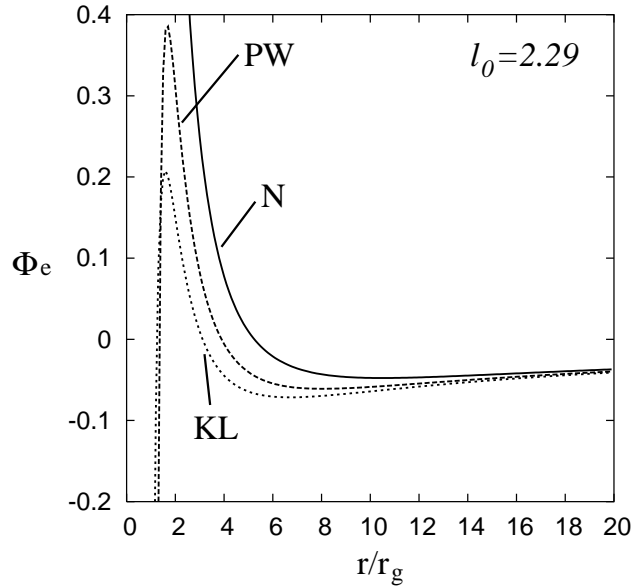}}
\caption{Effective potential wells for the Newtonian (N),
Pacyz\'{n}ski--Wiita (PW) and Klu\'{z}niak--Lee (KL) formulae given in
the text, in units of $c^{2}/2$. The value of angular momentum is
given in units of $r_{g}c$. Capture orbits are not possible in
Newtonian physics, but they appear qualitatively in the
pseudo--Newtonian potentials, as in General Relativity. The local
minimum in each curve gives the radius of the circular orbit possible
at the corresponding value of $\ell_{0}$.}
\label{fig:potentials}
\end{figure}

Figure~\ref{fig:potentials} shows the effective potential well as a
function of radius for the three forms mentioned here and the same
value of the orbital angular momentum.

\subsection{Fluid equilibrium configurations}\label{fluideq}

The fluid constituting the torus is located in the vicinity of the
mass $M$, which produces the potential $\Phi$. For simplicity we shall
neglect the contribution of the mass density of the fluid, $\rho$, to
the potential, and assume azimuthal symmetry, so that the problem can
be studied in cylindrical coordinates $(r,z)$. The gravitational pull
from the mass $M$ is countered by the fluid pressure gradients and the
centrifugal force, such that

\begin{equation}
\frac{1}{\rho} \nabla P = -\nabla \Phi_{\rm eff},\label{eq:hydroeq}
\end{equation}
where
\begin{equation}
\Phi_{\rm eff}= \Phi + \int \frac{\ell(r^\prime)^{2}}{2 r^{\prime 3}} dr
\end{equation}
is the effective potential and $\ell$ is the distribution of specific
angular momentum of the fluid, which depends only on the radial
coordinate (according to Von Zeipel's theorem, for a polytropic
equation of state the surfaces of constant angular velocity for a
rotating fluid are cylinders).

\begin{figure}
\resizebox{\hsize}{!}
{\includegraphics[]{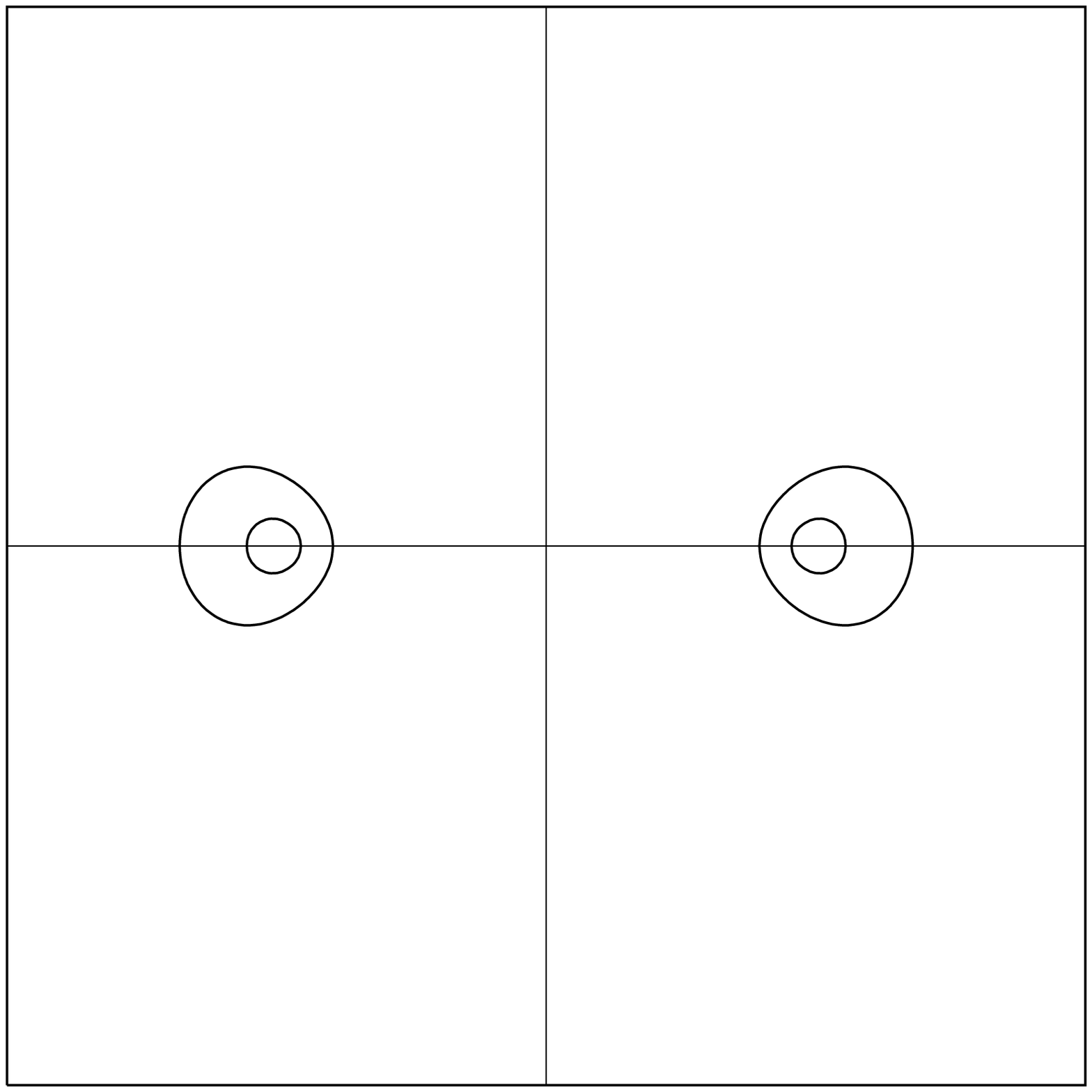}}
\resizebox{\hsize}{!}
{\includegraphics[]{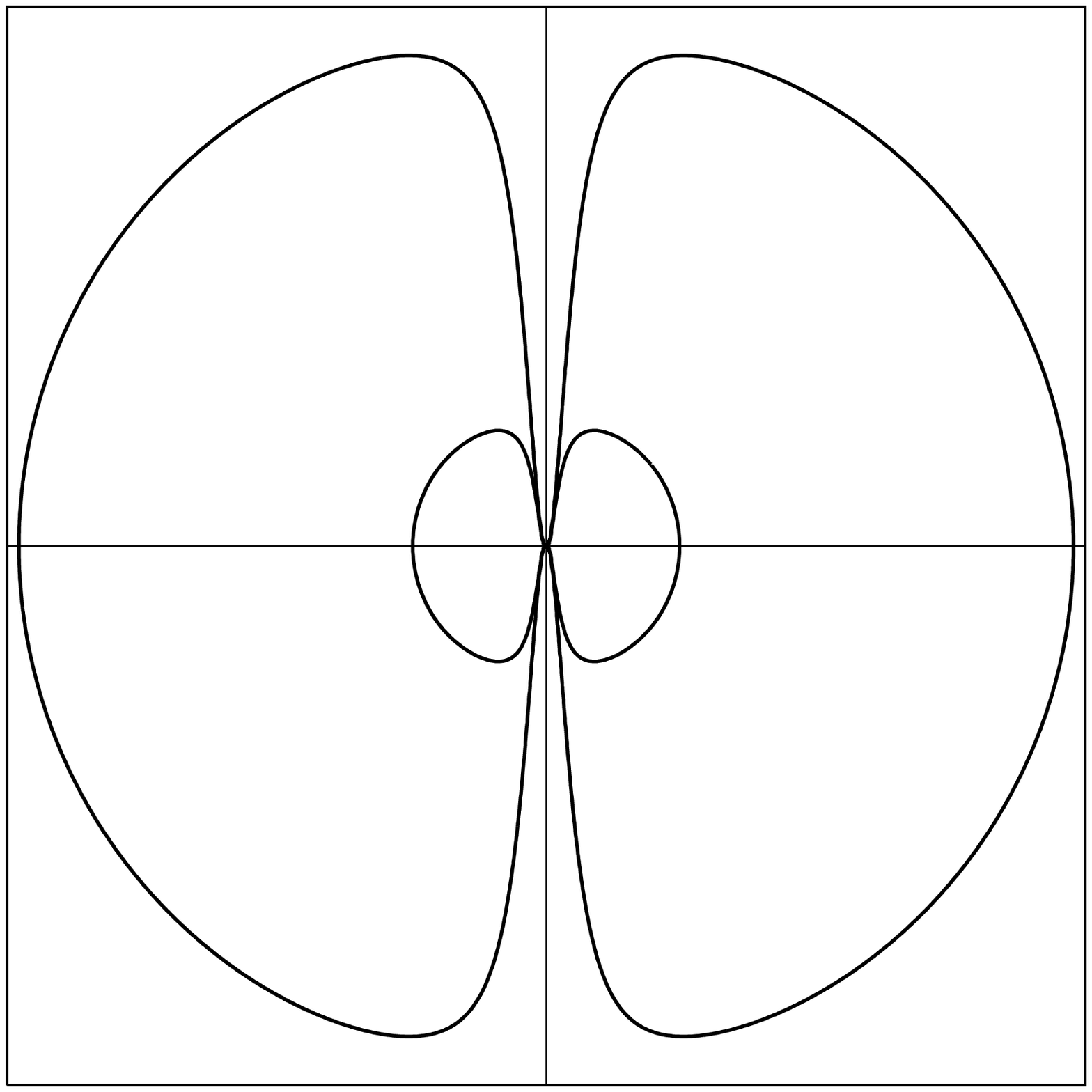}}
\caption{Meridional cross section of torus boundaries for a constant
distribution of angular momentum, $\ell$ and different values of
$\Phi_{0}$ in the Paczynski--Wiita potential. In the limit of a
slender torus (top) the meridional cross sections of the equipotential
surfaces become ellipses with axis ratio
$a_{r}/a_{z}=\nu_{r}/\nu_{z}$. In the limit of a thick torus (bottom),
the outer surface becomes spherical, while a narrow funnel develops
along the rotation axis. }
\label{fig:surfaces}
\end{figure}

Now, equation~(\ref{eq:hydroeq}) can be trivially integrated over $r$
if the fluid obeys a polytropic relation of the kind $P=K
\rho^{\gamma}$, with $K$ a constant and $\gamma$ the adiabatic index,
to give

\begin{equation}
\frac{\gamma}{\gamma-1} \frac{P}{\rho} + \Phi_{\rm eff} + \Phi_{0} = 0.
\end{equation}
The constant of integration $\Phi_{0}$ is related to the size of the
torus. The boundary of the torus is the surface over which the
pressure is zero, and it coincides with a particular equipotential
surface (see Figure~\ref{fig:surfaces}).

The fluid within a torus is in hydrostatic equilibrium, with the
pressure gradient balancing gravity vertically, and centrifugal forces
providing the bulk of the radial support against the gravitational
field. The circle of maximum density (or pressure) defines the center
of the torus at $r_{0}$. At this point, the pressure gradient
vanishes, and thus the rotation is exactly Keplerian, as for a free
particle. If the potential well is filled only slightly with gas, the
torus will be slender, and confined to regions close to the equatorial
plane. If the potential well is increasingly filled, the torus will
grow and deform until it overflows the inner Roche lobe and accretion
onto the central mass begins. This Roche lobe is analogous to that
which occurs in binary systems, except in this case it is an effect of
General Relativity (Abramowicz, Calvani \& Nobili 1983), specifically
of the existence of capture orbits for a finite value of angular
momentum (see Figure~\ref{fig:potentials}).

It is easiest to think of the rotation law through the distribution of
angular momentum, and convenient to write it as a power law, with
$\ell(r)=\ell_{0} (r/r_{0})^{\alpha}$. For Keplerian rotation in a
Newtonian potential we have, obviously, $\alpha=1/2$, while a constant
distribution of angular momentum, such as the one used to plot the
curves in Figures~\ref{fig:potentials} and \ref{fig:surfaces}, has
$\alpha=0$.  For sub--Keplerian rotation (i.e., $\alpha < 1/2$ in this
case), the torus is bound by a surface of zero pressure, becoming
infinite in the exact Keplerian limit (where one would have
pressureless dust in orbit). The overall shape and size of the torus
is thus determined by the degree to which the potential welll is
filled, and by the rotation law.

\section{Numerical Method and Initial Conditions}\label{methodICs}

\subsection{Numerical Method}\label{method}

We have used the SPH (Smooth Particle Hydrodynamics) numerical method
to perform our simulations (Monaghan 1992). This is a Lagrangian
formulation, which interpolates a given function $A(r)$, with a set of given
quantities $A(r')$ through a kernel function $W(r,h)$, using the
convolution integral:
\begin{equation}
A(r)=\int A(r')W(r-r')dr.
\end{equation}
We have implemented the prescriptions of  Monaghan \& Lattanzio 
(1985) for azimuthal symmetry, using the  kernel:
\begin{equation}
W(r,h)= \frac{\sigma}{h^{\nu}} \left\{ \begin{array}{l l}
         1-3\left(\frac{r}{h}\right)^{2}/2+3\left(\frac{r}{h}\right)^{3}/4 
         & \mbox{if $0
         \leq \frac{r}{h} \leq 1$;} \\ & \\ \left(2-\frac{r}{h}\right)^{3}/4 &
         \mbox{if $1 \leq \frac{r}{h} \leq 2$;} \\ & \\ 0 & \mbox{2 $\leq
         \frac{r}{h}$.}
         \end{array}
         \right.
\end{equation}
Here $h$ represents the smoothing length, and is
comparable to the typical separation between fluid elements (it
essentially gives the spatial resolution of the calculation), and $r$
is the radial coordinate.  In two dimensions, $\nu=2$ and
$\sigma=10/7\pi$. The gas is modeled as an inviscid fluid, and so 
the Navier--Stokes equations take the form:
\begin{equation}
\frac{dv_{r}}{dt}=- \frac{1}{\rho}\frac{\partial P}{\partial r}
-\frac{GM_{BH}r}{R(R-r_{g})^{2}}+r \Omega^{2}
+\left(\frac{dv_{r}}{dt}\right)_{\rm art},
\end{equation}
\begin{equation}
\frac{dv_{z}}{dt}=- \frac{1}{\rho}\frac{\partial P}{\partial z}
-\frac{GM_{BH}z}{R(R-r_{g})^{2}}+
\left(\frac{dv_{z}}{dt}\right)_{\rm art},
\end{equation}
where $R=\sqrt{r^2+z^2}$ is the distance to the central mass $M$.  The
sub--index \emph{art} indicates the artificial viscosity terms, which
is used to account for the presence of shocks and to avoid 
interpenetration of the fluid elements.

The energy equation is:
\begin{equation}
\frac{du}{dt}=- \left(\frac{P}{\rho}\right)\nabla \cdot \vec{v}+
 \left( T \frac{ds}{dt} \right)_{\rm art}
\end{equation}
where $u$ is the internal energy per unit mass. No external (i.e.,
radiative) losses are assumed, and thus the only change in $u$ arises
from work done. When discretized over a finite number of fluid
elements, often called particles in SPH, the convolution integral
becomes a sum over all elements.

We have used the prescription given by Balsara (1995) for the
artificial viscosity, which reduces the artificial shearing
stress. The explicit forms for the acceleration and the energy
dissipation due to artificial viscosity for the $i-eth$ SPH fluid element are:
\begin{equation}
\left(\frac{d \vec{v}}{d t}\right)_{i,\rm art}=
- \sum_{j \neq i}m_{j} \Pi_{ij}\nabla_{i}W_{ij},
\end{equation}
and
\begin{equation}
\left(T \frac{ds}{dt}\right)_{i,\rm art}= \frac{1}{2} \sum_{j \neq i}m_{j}
\Pi_{ij}(\vec{v_{i}}-\vec{v_{j}}) \cdot \nabla_{i}W_{ij}.
\end{equation}
where $\Pi$ is defined by (see, e.g., Lee and Ramirez-Ruiz 2002)
\begin{equation}
\Pi_{ij}=\left(\frac{P_{i}}{\rho_i^{2}}+\frac{P_{j}}{\rho_{j}^{2}}\right)=
(-\alpha_{b}\mu_{ij}+\beta_{b}\mu_{ij}^{2}),
\end{equation}
\begin{equation}
\mu_{ij}= \left\{ \begin{array}{l l} \frac{(\vec{v}_{i}-\vec{v}_{j}) \cdot
(\vec{r}_{i}-\vec{r}_{j})}
{h_{ij}(|\vec{r}_{i}-\vec{r}_{j}|^{2}/h_{ij}^{2})+\eta^{2}}
\frac{(f_{i}+f_{j})}{2c_{ij}}
& \mbox{if $\vec{v}_{ij} \cdot \vec{r}_{ij} < 0$;}
\\ 0 & \mbox{if $\vec{v}_{ij} \cdot \vec{r}_{ij}
\geq 0$;}
         \end{array}
         \right.
\end{equation}
The function $f_{i}$ is defined by:
\begin{equation}
f_{i}=\frac{|\nabla \cdot \vec{v}|_{i}}
{|\nabla \cdot \vec{v}|_{i}+|\nabla \times \vec{v}|_{i} + \eta'c_{i}/h_{i}},
\end{equation}
and $\eta=10^{-2}$.  The sound speed and smoothing length of
each element are denoted by $c_{i}$ and $h_{i}$ respectively, and the
factor $\eta^{\prime} \simeq 10^{-4}$ in the denominator prevents
numerical divergences.  $\alpha_{b}=\beta_{b}=\gamma/2$ are constants of
order unity and $\gamma$ is the polytropic index from the equation of
state. This form of the artificial viscosity supresses the shearing
stresses when the compression in the fluid is low and the vorticity is
high, $|\nabla \cdot \vec{v}| \ll |\nabla \times \vec{v}|$, but
remains in effect if the compression dominates in the flow $|\nabla
\cdot \vec{v}| \gg |\nabla \times \vec{v}|$.

\subsection{Initial Conditions and Applied Perturbations}\label{ICspert}

To construct a particular fluid torus, one needs only to specify the
equation of state, the distribution and absolute value of angular
momentum, and the degree to which the potential well is filled
(through the constant $\Phi_{0}$). We restrict ourselves here to
systems in which $\ell(r)=\ell_{0}=$~cst. Thus the actual value of
angular momentum fixes the position of the center of the torus,
$r_{0}$, as defined in \S~\ref{fluideq}.  Numerically, this is done by
means of a Monte Carlo technique, by distributing fluid elements over
the prescribed volume according to the analytical density profile, and
subsequently relaxing them with artificial damping included in the
equations of motion. The rotation law is strictly enforced during this
procedure, and the internal energy is fixed by the adiabatic relation
so that one has complete control over the initial condition. We have
verified that our initial conditions are indeed in equilibrium by
evolving them explicitly in time without any applied perturbations. No
global evolution is observed during these computations.

We have applied two fundamentally different types of perturbations to
initial models: impulsive and periodic. In the first, an additional
velocity field is imposed on the initial condition at $t=0$, and the
system is evolved without additional perturbations for many (typically
one hundred) dynamical times. In the second, an additional, periodic
term is added to the equations of motion to produce an
acceleration. This can be applied only in the radial direction, or
vertical, or both.

Finally, one can hold $\ell_{0}$ constant during a calculation, or
vary it slowly (i.e., over many dynamical times). In the first case,
the torus will remain essentially in the same radial region,
oscillating due to the applied perturbation. In the second, in
addition to these oscillations it will either move inward or outward,
depending on whether $\ell_{0}$ decreases or increases. We have
considered this possibility in view of the fact that gas within an
accretion disk will, in reality, drift radially as angular momentum is
transported by viscous stresses. 

The temporal profile of the applied forcing can be varied. We have
used single--frequency sinusoids as well as narrow pulse--like
functions with repetition time $T_{s}=1/\nu_{s}$. This can be thought
of as the rotation period of the central mass, which affects the
accretion disk through some unspecified mechanism (e.g., the pulsar
magnetic field, asymmetries in the quadrupole moment or other
effects).

\section{Forced Oscillations}\label{forcedosc}

We will hereafter restrict the presentation to the case of slender
tori, where their radial and vertical extent, $L$, is small compared
to the radial position of their center, i.e., $L \ll r_{0}$. The thin
annulus can then be viewed as a small section of the entire accretion
structure surrounding the compact object. The main difficulty with
this picture, and in relating it to real systems, lies in the fact
that the artifical zero--pressure boundaries obviously make wave
propagation into the exterior of this torus impossible. Mode leaking
is certainly an important issue (Fragile 2005), but for now we address
only closed systems. The dynamical response of thick tori under global
impulsive perturbations has been addresssed in a series of papers by
Rezzolla and collaborators (Zanotti et al. 2003; Rezzolla et
al. 2003a,b; Montero et al. 2004; Zanotti et al. 2005), while
global and localized periodic perturbations have been considered by
Rubio--Herrera \& Lee (2005a,b).

As a result of the induced forcing, the torus experiences
small--amplitude oscillations, which can be measured in various
ways. One of these is to simply consider the position of its center,
$(r_{0},z_{0})$, defined, as above, as the location of maximum
density, as a function of time. The corresponding time series are
complicated, as they prominently display the perturbing frequency
$\nu_{s}$ and the local epicyclic frequency for small radial and
vertical oscillations, $\nu_{r}$ and $\nu_{z}$ respectively.  There
are however, combination frequencies and cross--over phenomena, so
that the whole behavior is best analyzed by performing Fourier
transforms of $r_{0}(t)$ and $z_{0}(t)$.

\subsection{Radial and Vertical Mode Coupling}\label{rzcoupling}

\begin{figure}
\resizebox{\hsize}{!}
{\includegraphics[]{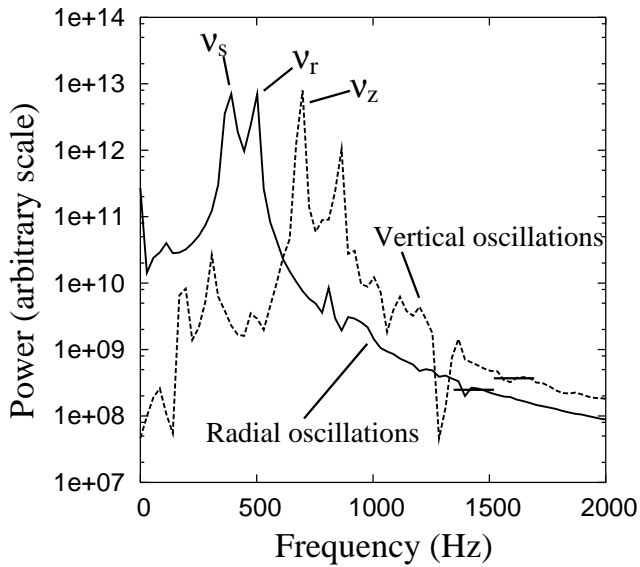}}
\caption{Fourier transforms of the radial (solid) and vertical
(dashed) oscillations of the center of a slender torus with constant
specific angular momentum when perturbed periodically at frequency
$\nu_{s}=400$~Hz in a pseudo--Newtonian potential. The local values of
the radial and vertical epicyclic frequencies are $\nu_{r}=500$~Hz and
$\nu_{z}=700$~Hz, respectively. Even though the perturbation is purely
radial, vertical oscillations are excited because of pressure
coupling. The power was re--scaled along the vertical axis for
illustrative purposes. In reality the power in vertical motions is
much smaller than in the radial ones.}
\label{fig:fftfixedr} 
\end{figure}

\begin{figure}
\resizebox{\hsize}{!}
{\includegraphics[]{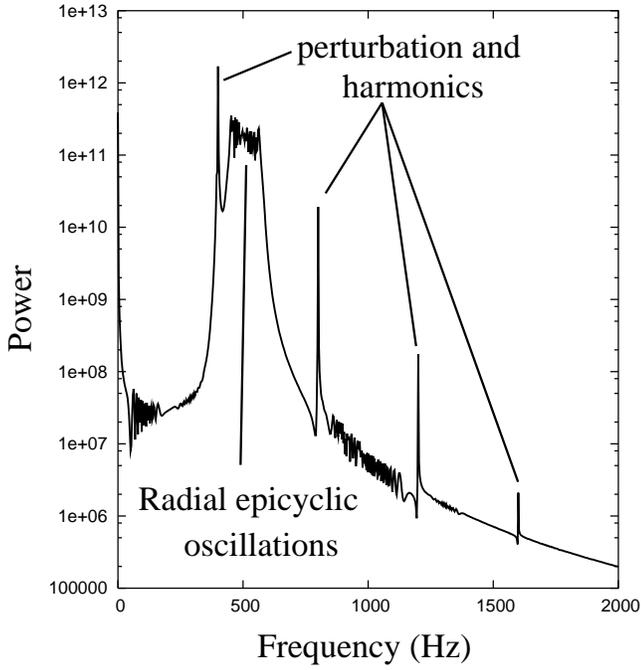}}
\resizebox{\hsize}{!}
{\includegraphics[]{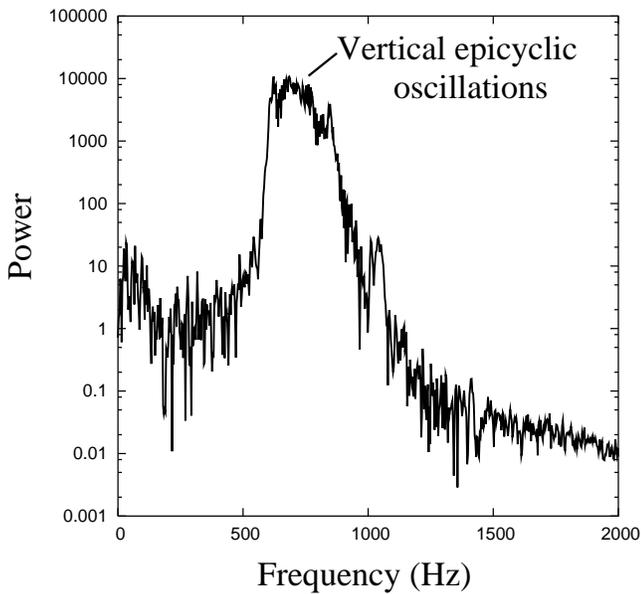}}
\caption{Fourier transforms of the radial (top) and vertical (bottom)
oscillations of the center of a radially drifting slender torus when
perturbed periodically at frequency $\nu_{s}=400$~Hz in a
pseudo--Newtonian potential. The outer (initial) and inner (final)
central radii are $r_{0}=6.7 r_{g}$ and $r_{0}=5.35 r_{g}$
respectively.}
\label{fig:fftcmfullt} 
\end{figure}

\begin{figure}
\resizebox{\hsize}{!}
{\includegraphics[]{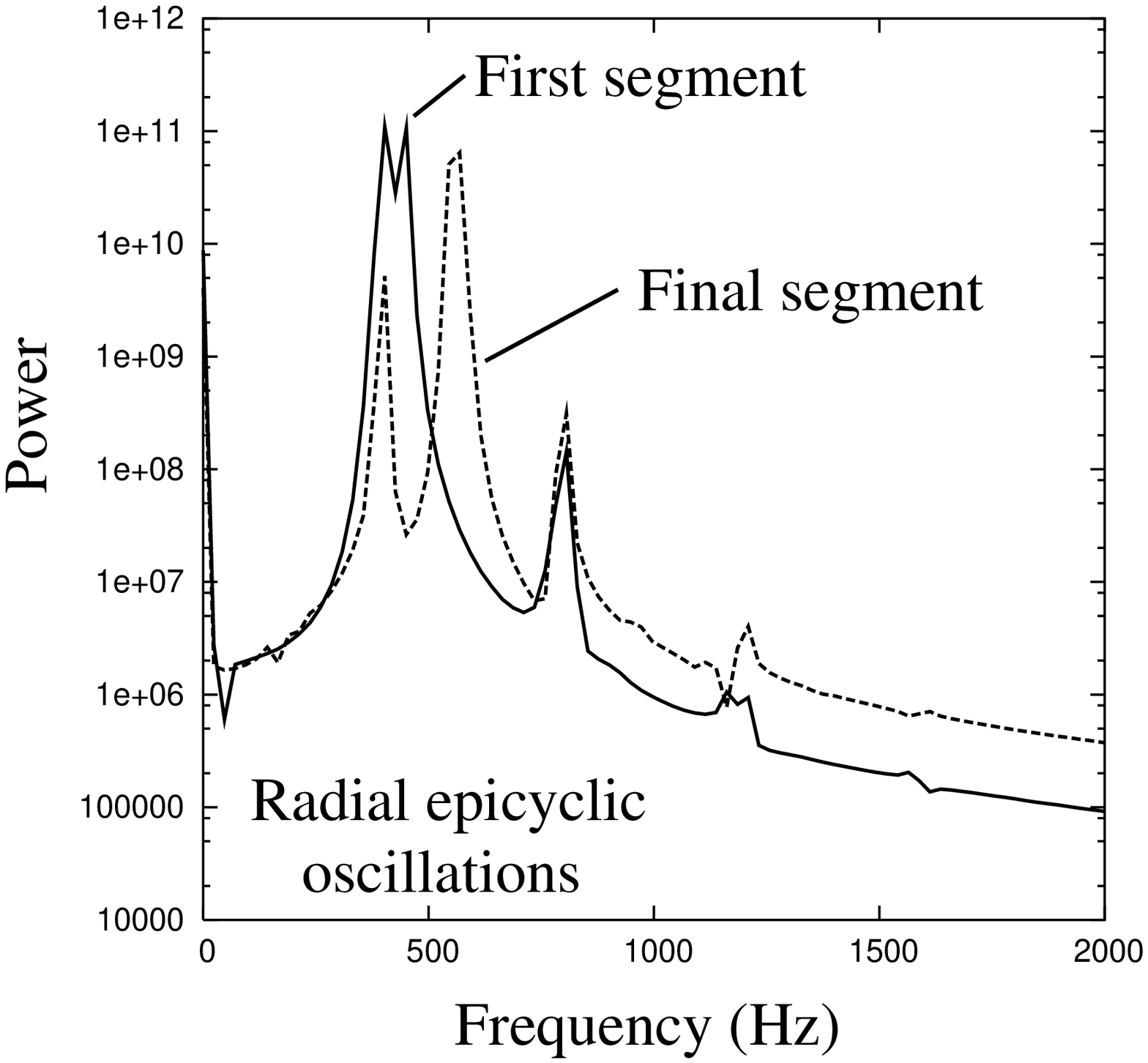}}
\resizebox{\hsize}{!}
{\includegraphics[]{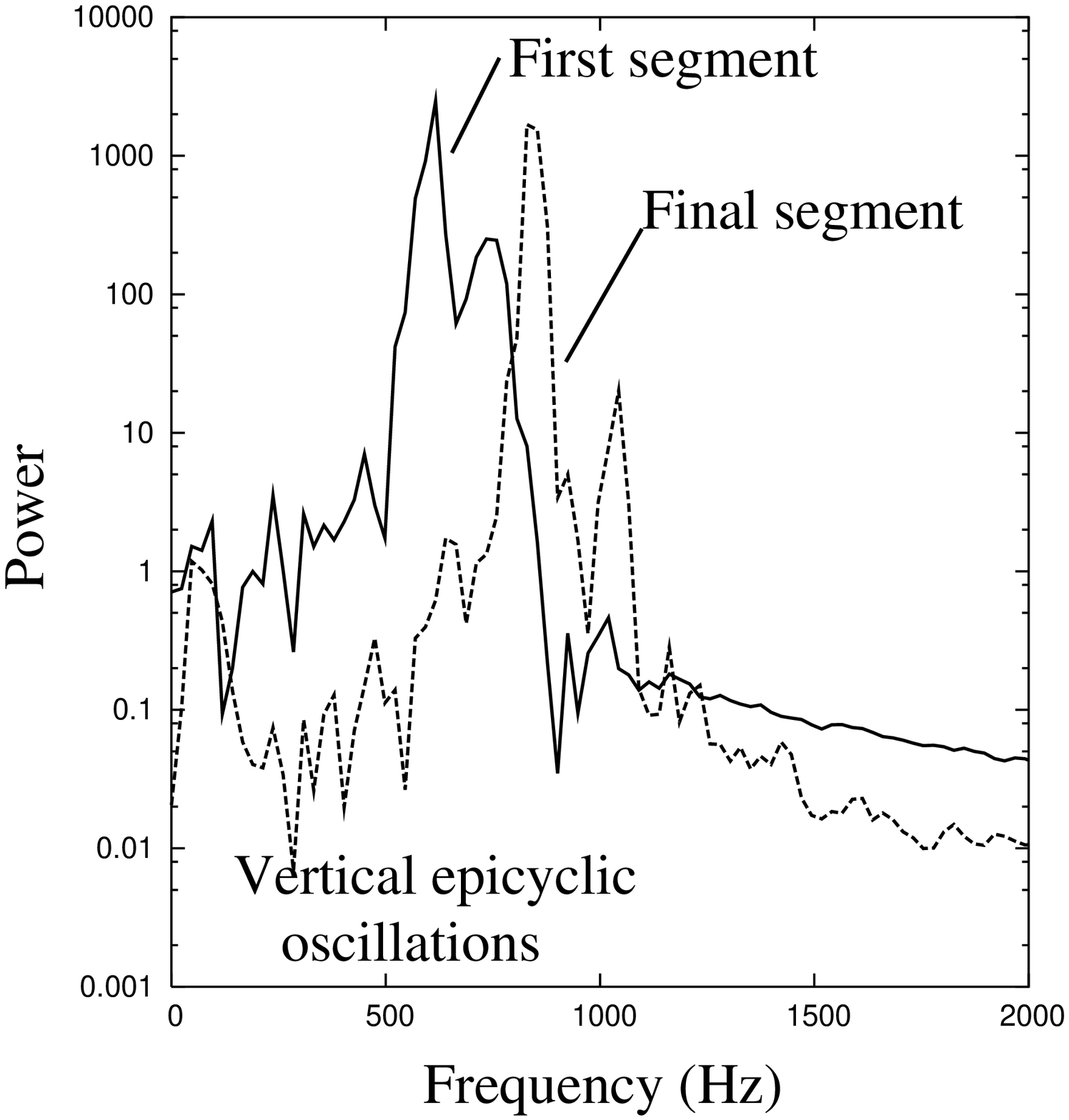}}
\caption{Fourier transforms of the radial (top) and vertical (bottom)
oscillations of the center of a radially drifting slender torus when
perturbed periodically at frequency $\nu_{s}=400$~Hz in a
pseudo--Newtonian potential. Only the first and last segments of the
calculation (1/8 of the total) are shown.}
\label{fig:fftcm18} 
\end{figure}

For calculations where the angular momentum does not vary in time,
since it is also constant in space, there are no mechanisms for its
transport, and the perturbation does not produce any torques, the
fluid must remain close to its equilibrium radius
$r_{0}$. Figure~\ref{fig:fftfixedr} shows the Fourier transform of
$r_{0}$ and $z_{0}$ for such a calculation, assuming the potential
$\Phi_{\rm KL}$, a central mass $M=1.4$~M$_{\odot}$,
$r_{0}=6.1$~r$_{g}$ and a purely radial, sinusoidal perturbation at
frequency $\nu_{s}=400$~Hz. The corresponding values of the local
radial and vertical epicyclic frequencies are $\nu_{r}=500$~Hz and
$\nu_{z}=700$~Hz. The power spectrum of radial motions clearly shows
the perturbation frequency, $\nu_{s}$, and the radial epicyclic
frequency, $\nu_{r}$. Likewise, the power spectrum of vertical motions
exhibits the vertical epicyclic frequency $\nu_{z}$. In this
particular case, the value of $r_{0}$ and $\nu_{s}$ is such that the
difference between the two epicyclic frequencies is equal to half the
spin frequency, i.e., $\nu_{z}-\nu_{r}=\nu_{s}/2$. There is thus clear
evidence for mode coupling, through pressure, since the perturbation
was initially only applied in the radial direction. Beats between the
various frequencies are also present. In Figure~\ref{fig:fftfixedr},
the power in vertical motions shows peaks at $\nu_{z}-\nu_{s}$,
$\nu_{r}$, and $\nu_{s}+\nu_{r}$.  The very mode responsible for the
coupling between the radial and vertical oscillations is weakly
visible at $\nu_{z}-\nu_{r}=200$~Hz.

\begin{figure}
\resizebox{\hsize}{!}
{\includegraphics[]{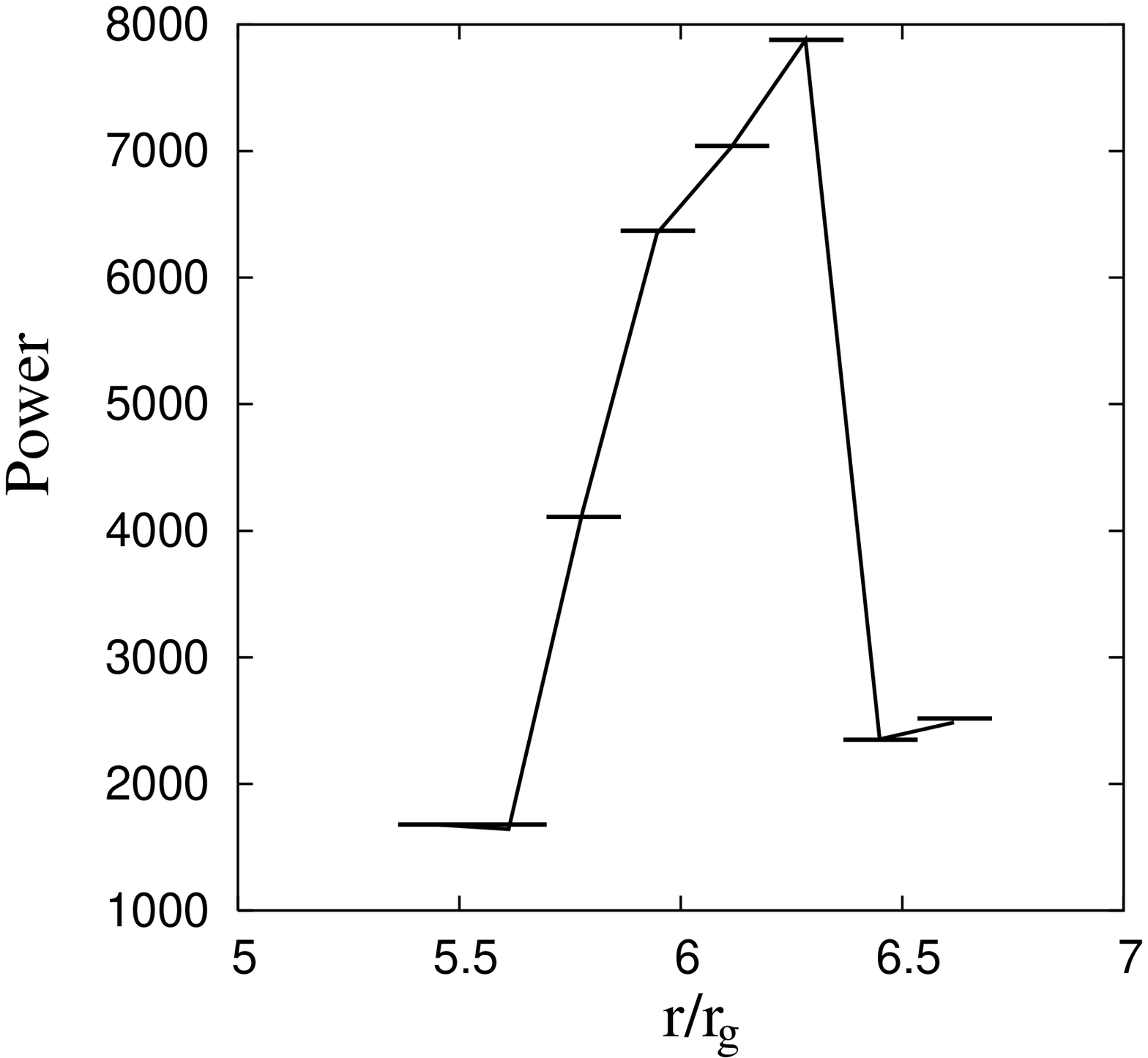}}
\caption{Power in vertical motions at the local value of $\nu_{z}$ as
a function of the average central position of the torus, $r_{0}$ for
the calculation with a radially drifting (inwards) perturbed
torus. The calculation was divided into eight segments of equal
duration. During each segment the torus drifted by $0.16 r_{g}$ in the
radial direction (shown by the horizontal bars). The strongest
response occurs at $r_{0}\approx 6.15 r_{g}$, where
$\nu_{z}-\nu_{r}=\nu_{s}/2$. Note how the curve is not symmetrical with
respect to the point of maximum power. See the text for details.}
\label{fig:zpowervsr} 
\end{figure}

A calculation with varying angular momentum (see \S~\ref{ICspert})
where the torus drifts inward is even more revealing. Starting at
$r_{0}=6.7 r_{g}$, and terminating at $r_{0}=5.35 r_{g}$ over two
hundred initial orbital periods, the radial and vertical oscillations
of the center of the torus occur at a range of frequencies, covering
values between the extremes in epiclycic frequencies $\nu_{r}$ and
$\nu_{z}$.  This choice of parameters implies that, with
$\nu_{s}=400$~Hz, the calculation begins with $\nu_{z}-\nu_{r} <
\nu_{s}/2$ and ends with $\nu_{z} - \nu_{r} > \nu_{s}/2$.

A power spectrum of the radial oscillations of $r_{0}$ during the a
complete simulation using a pulse--like perturbation shows several
features (see Figure~\ref{fig:fftcmfullt}). The perturbation at
$\nu_{s}$ is clearly visible, as are its three harmonics (it is not a
pure sine wave, see Figure~3 in Lee, Abramowicz \& Klu\'{z}niak
2004). A broad, flat--topped peak is also apparent, covering the range
$ 440 {\rm Hz} < \nu_{r} < 580 {\rm Hz}$. These are simply the local
values of $\nu_{r}$ at the inner and outer radii, respectively.
Harmonics of the radial epicyclic frequency, as well as a beat with
$\nu_{s}$ are also visible, at greatly reduced power. The
corresponding power spectrum for the oscillations of $z_{0}$ shows a
similar broad peak, for $ 590 {\rm Hz} <\nu_{z} < 880 {\rm Hz}$, as
expected (see Figure~\ref{fig:fftcmfullt}).

To explore the behavior of the oscillations in more detail, we have
split the time series of $r_{0}(t)$ and $z_{0}(t)$ into eight equal
segments, each thus covering one eighth of the simulation. During this
time, the torus is in a narrow radial window, and we may say that its
position is nearly constant (except for the oscillations induced by
the external forcing). We then extract the individual Fourier
transforms for each segment. The results for the radial oscillations
are shown in Figure~\ref{fig:fftcm18}, for the first and last time
segments (they are all qualitatively similar). As expected, each one
shows the perturbation and its harmonics, as well as a narrow peak
centered at the value of $\nu_{r}$ corresponding to the average
position of the circle of maximum pressure during the time
interval. The power spectrum of the corresponding vertical
oscillations is shown in Figure~\ref{fig:fftcm18}. There is a strong
peak at $\nu_{z}$, the amplitude of which varies greatly as the torus
drifts radially. In Figure~\ref{fig:zpowervsr} we show the power in
this peak as a function of the average position of the center of the
torus, $r_{0}$. Two facts are immediately clear. First, the intensity
of the coupling between radial and vertical modes is a function of the
external perturbation. Second, this is a non--linear coupling, because
the interaction is through a sub--harmonic of the pertrubation,
$\nu_{s}/2$, and not with the perturbation itself or its higher
harmonics. This fact points clearly to rich non--linear behavior in
these systems, and, we believe, has been directly seen in accretion
disks in LMXBs (Klu\'{z}niak et al. 2004; Lee, Abramowicz \&
Klu\'{z}niak 2004), in particular in SAXJ1808.4-3658 (Wijnands et
al. 2003). 

If the coupling between modes is due to some interaction with the
external perturbation, and this excites a resonance, one would naively
expect the consequences to display resonance--like behavior. In
particular, the resonant window should be narrow, and if the
corresponding condition is not met (e.g., above
$\nu_{z}-\nu_{r}=\nu_{s}/2$), then no coupling should occur. However,
this is not what we observe in the calculation we have just
described. As the torus drifts inward, the power in vertical motions
rises sharply as the resonance is approached, {\em but remains high
even after it has been passed}. The corresponding range in frequencies
for $\nu_{z}$ spans more than 100~Hz. This appears to indicate that
once excited, the modes have locked (Klu\'{z}niak et al. 2004) and
continue to be excited for some time. The curve in
Figure~\ref{fig:zpowervsr} is reminiscent in shape of the amplitudes
of meridional oscillations for slightly perturbed orbits in nearly
geodesic motion, as computed by Abramowicz et al. (2003, see their
Figure~2). In that case a generic coupling between radial and polar
oscillators was assumed. In Neutron Star sources, the frequencies of
the kHz QPOs drift over large ranges, sometimes hundreds of Hz. Mode
locking following resonant excitation could in principle be
responsible for this behavior.

\subsection{Additional Vertical Modes}\label{morevertical}

Inspection of Figures~\ref{fig:fftfixedr} and \ref{fig:fftcm18}
reveals that the power spectrum of vertical oscillations has a
secondary peak at frequency $\nu_{u} > \nu_{z}$. The power in this
oscillation is variable as the torus drifts radially, but it appears
to be continously excited. In some instances it seems to increase in
power when $\nu_{z} - \nu_{r} \approx \nu_s/2$, but its presence is
not limited to the corresponding radial window.

Thus it would seem again, in agreement with our previous results, that
a non--linear coupling between the radial and vertical modes transfers
energy to the latter. Further, a second oscillation at frequency
$\nu_{2} \approx \nu_{z} + \nu_{s}/2$ is excited within the same
vertical motions, with power comparable to that at frequency
$\nu_{z}$. Previously, when we examined peaks in the power spectrum at
two different frequencies, they occurred in different variables (e.g.,
radial vs. vertical) and thus it was impossible to directly compare
their amplitudes (as the strength of the real coupling is unknown, and
the applied perturbation is arbitrary). Here we now show for the first
time evidence for similar amplitudes in different modes of the same
general motions.

\section{Conclusions and Directions for Future Work}

The oscillations of fluid confined by gravity and/or centrifugal
forces can lead to rich behavior under a variety of astrophysical
scenarios (we have focused on QPOs in X--ray binaries, the
observations of solar oscillations provide another, much closer
example). Under quite general considerations we have shown that
simple, acoustic modes which have their origin in free particle
oscillations are modified by the hydrodynamics and can couple to one
another in non--linear fashion. The locking of excited modes could
explain the fact that the observed QPO frequencies drift over
considerable ranges while still arising from resonant interactions. We
believe that their signature is present in the observations of
accretion flows in strong gravitational fields, and will allow for the
measurement of important parameters related to the central, compact
object.

Clearly, meaningful constraints will be derived only through much more
detailed work. For example, it is not clear how such oscillations will
actually modify the X--ray lightcurve (see, e.g., Schnittman 2005 for
work along these lines). In addition, the strong gravitational field
needs to be modeled in full General Relativity, with consideration of
the self--gravity of the disk and detailed thermodynamics and
radiation transport. 

\acknowledgements

It is a pleasure to thank M.~A. Abramowicz and W. Klu\'{z}niak for a
continous and enjoyable collaboration. Financial support for this work
was provided in part by CONACyT (36632E) and NORDITA, whose
hospitality is gratefully acknowledged.

\end{document}